\documentclass[aps,prb,reprint,amsmath,amssymb,superscriptaddress,citeautoscript,longbibliography,floatfix]{revtex4-2}

\newcommand{\angstrom}{\textup{\AA}}
\usepackage[allow-number-unit-breaks]{siunitx}
\usepackage{graphicx}
\usepackage{dcolumn}
\usepackage{bm}
\usepackage{color}
\usepackage{bbding}
\usepackage{pifont}
\newcommand{\xmark}{\ding{55}}
\usepackage{mathrsfs}
\usepackage{amsfonts}
\usepackage{multirow}
\usepackage{array}
\usepackage{natbib}
\usepackage{epsf}
\usepackage{epsfig}
\usepackage{epstopdf}
\usepackage{hyperref}
\hypersetup{
    colorlinks=true,
    linkcolor=blue,
    filecolor=cyan,      
    citecolor=blue,
    urlcolor=blue,
    }

\begin{document}

\title{Spin-chirality-driven nonrelativistic Edelstein effects in two-dimensional antiferromagnets}

\author{Hao Zuo}
\affiliation{Center for Alloy Innovation and Design, State Key Laboratory for Mechanical Behavior of Materials, Xi'an Jiaotong University, Xi'an, 710049, China}

\author{Xiaoyin Li}
\affiliation{Center for Alloy Innovation and Design, State Key Laboratory for Mechanical Behavior of Materials, Xi'an Jiaotong University, Xi'an, 710049, China}

\author{Jian Zhou}\email{jianzhou@xjtu.edu.cn}
\affiliation{Center for Alloy Innovation and Design, State Key Laboratory for Mechanical Behavior of Materials, Xi'an Jiaotong University, Xi'an, 710049, China}

\begin{abstract}
Charge current-induced magnetic moment accumulation--Edelstein effect--has been extensively attracting attention for its promising applications in spintronics. While most prior works focus on the spin-orbit coupling (SOC) induced Edelstein responses that rely on the presence of heavy elements, the nonrelativistic Edelstein effect (in the absence of SOC) that could be applied in a broader material family has been largely unexplored. Here, we perform a combined group-theoretical and \textit{ab initio} numerical simulation study to show that vector spin chirality could serve as an effective control parameter of nonrelativistic Edelstein responses in antiferromagnetic system. In addition to spin degree of freedom, we also explore the orbital angular momentum contributions to current-induced magnetic moments (dubbed orbital Edelstein effect), which obey distinct symmetry constraints from the spin counterpart. Microscopically, vector spin chirality $\kappa$ gives rise to electronic and Zeeman-like band-geometric quantities, such as the anomalous spin/orbital polarizability and the Berry connection polarizability, which govern the nonrelativistic Edelstein responses. Our work identifies vector spin chirality as a key magnetic order parameter to enable and tune nonrelativistic Edelstein effects, and uncovers a new route toward electrically controlling magnetization without relying on SOC effect.
\end{abstract}

\maketitle
\section{Introduction}
The electrical manipulation of magnetic order is one of the central objectives of modern spintronic technology \cite{Baltz2018}. While the spin transfer torque (STT) \cite{Slonczewski1996,Berger1996} occurs with a spin polarized current passing through the magnetic material, another mechanism--spin orbit torque (SOT)--usually relies on current flowing in the adjacent nonmagnetic heavy metal conductor. In this approach, the spin torque is usually generated through significant spin-orbit coupling (SOC)  with Edelstein effect and/or inverse spin galvanic effect \cite{Manchon2019,Haney10PhysRevLett.105.126602,Freimuth14PhysRevB.90.174423}. Compared to the STT mechanism, the SOT strategy is particularly attractive for lower power consumption, faster magnetization dynamics, and the absence of a current directly passing through the magnetic layer. The Edelstein effect describes a charge current drives the electronic out-of-equilibrium distribution and furthermore produces spin accumulations due to spin-momentum locking in the Brillouin zone (BZ). It then exerts a torque on the local magnetic moments \cite{Edelstein1990,Ivchenko1978,Aronov1989,Zelezny2014,Johansson2024,Tenzin2023,Roy2022}, realizing controlling of magnetization using electric field. According to symmetry argument, as an inversion symmetry operator $\bar{E}$ flips the electric field component $\bm E$, the linear Edelstein effect (LEE) requires inversion symmetry breaking in its host system \cite{Johansson2024,Tenzin2023}, which limits the material selection for its future developments.

To overcome this symmetry constraint, recent works have been devoted to nonlinear (i.e., the second order) Edelstein effect (NLEE) \cite{Xiao2022PRL129,Xiao2023PRL130,Baek2024,WTe2NL,Xu21PRB,Sarkar_2025NJPae1868}. Here, since $\bar{E}$ does not flip the second order electric field $E^2$, it is symmetry-allowed by inversion symmetry, suggesting that the NLEE could emerge even in centrosymmetric crystals. This largely expands the range of candidate materials for electrically generated spin polarization. Note that similar mechanisms have also been proposed for the optical control of magnetization in a single phase material, with finite light frequencies included in the response theory \cite{Xu21PRB,zhou2025contrasting,zhou2022photo}.  From a microscopic perspective, both LEE and NLEE responses can be partitioned into time-reversal ($\mathcal{T}$)-odd and $\mathcal{T}$-even contributions, which originate from distinct band-geometric quantities. In detail, the intrinsic $\mathcal{T}$-odd LEE mainly arises from the Zeeman Berry curvature (see below) of the Fermi sea contributions, while its $\mathcal{T}$-even term is from the Zeeman quantum metric at the Fermi surface. As for the NLEE, its time-reversal ($\mathcal{T}$)-odd part is mainly governed by the Berry connection polarizability (BCP) ~\cite{Xiao2022PRL129,WTe2NL,Ye2025WTe2AHE}, while the $\mathcal{T}$-even term is described by the momentum space dipole of the anomalous spin polarizability (ASP) ~\cite{Xiao2023PRL130}.
In the prior works that mainly focus on nonmagnetic and collinear magnetic systems, both LEE and NLEE requires relativistic SOC effects to yield finite magnetoelectric interactions \cite{Trama2024}. Hence, it is usually required to adopt material platforms containing heavy elements with strong SOC to realize sizable Edelstein effects.

Recent years have witnessed rapidly growing interest in the nonrelativistic (NR, i.e., the SOC-free limit) effect on the spin-momentum locking and its related effects \cite{GonzalezHernandez2024,Hu2025,Chakraborty2025}. For instance, the exchange field induced spin splitting has been theoretically predicted and experimentally demonstrated in several unconventional antiferromagnetic (AFM) systems, which exhibit $p$-, $d$-, $f$-, $h$-, and $i$-wave spin textures in momentum space \cite{Hayami2020,Smejkal2022AM,Smejkal2022Landscape,Yuan21PRM,Yuan23NC,Krempasky24Nature,Fedchenko24SA,Lee24PRL}. The resulting spin splitting can reach $0.1-1~\mathrm{eV}$, exceeding conventional SOC-induced spin splitting ($\sim 1-10~\rm meV$) by two to three orders of magnitude. Such a NR-limit is attractive especially in noncollinear magnetic systems, where responses traditionally associated with SOC, such as the anomalous Hall effect and magneto-optical Kerr effect, can persist even in the absence of SOC \cite{Chen2014PhysRevLett.112.017205,Feng2020NC,Kipp2021ChiralHall,Zhu2025MagneticGeometry}. From the perspective of spin-group theory \cite{Liu2022SPG,Xiao2024SSG}, these emergent responses originate from the reduction of spin symmetry from collinear to noncollinear magnetic orders. Consequently, noncollinear magnets have recently emerged as a fertile platform for exploring unconventional electric and spin transport, and electromagnetic interplay in the NR regime. Despite of these efforts, it remains largely underexplored how the LEE and NLEE can be triggered and continuously manipulated in the NR limit. In particular, we note that the magnetic moments actually compose two different sources, namely, spin angular momentum (SAM) and orbital angular momentum (OAM). With finite SOC effect, both of them share the same symmetry constraints and can be treated simultaneously. By contrast, when SOC is turned off, their responses would differ substantially and should be discussed separately. The decoupling of SAM and OAM in the NR regime would lead to novel physical responses and offer a pathway towards probing the underlying electronic and Zeeman-like band geometries through nonrelativistic Edelstein effects. Therefore, uncovering the microscopic origins and tunability of these responses in the NR regime represents an intriguing direction for exploration. 

\begin{figure}[t]
\centering
\includegraphics[width=0.9\columnwidth]{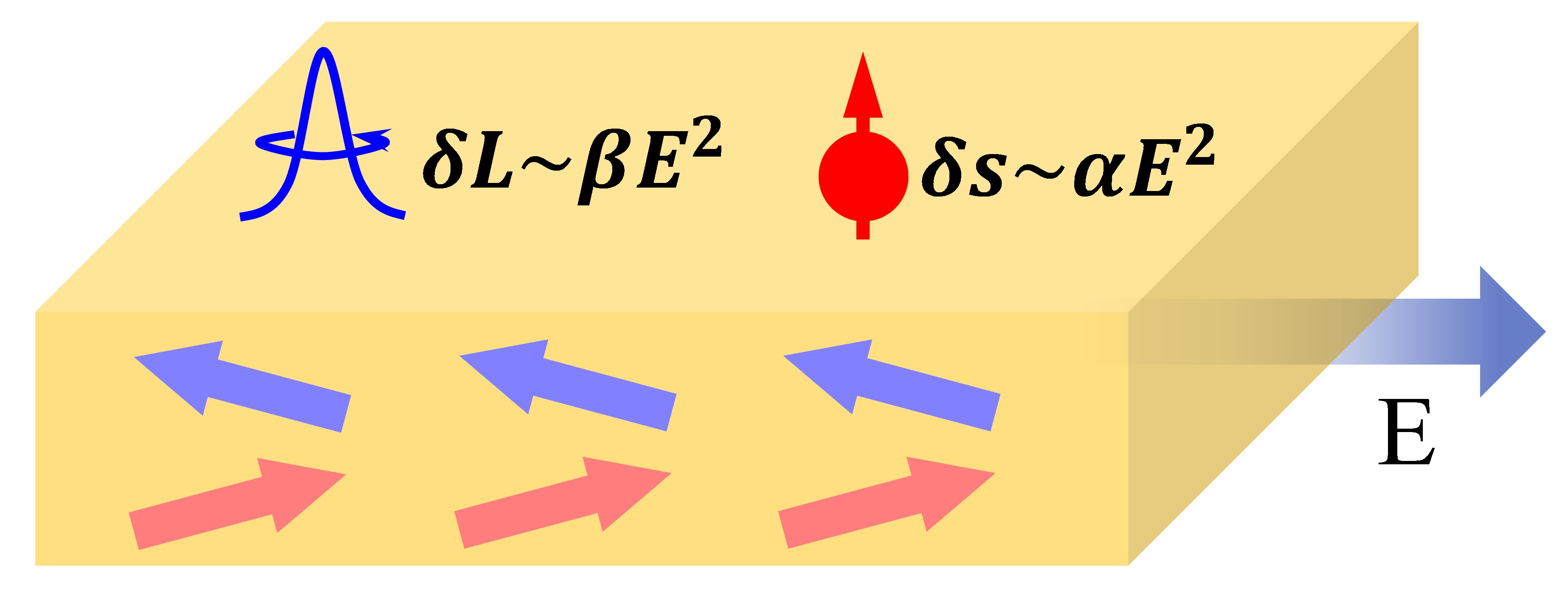}
\caption{Schematic plot of Edelstein effect in canted spin configuration. The in-plane electric field could induce spin and orbital magnetic moment accumulation, in the SOC-free limit.}
\label{fig:schematic}
\end{figure}

In this work, we address the above questions using a quasi-two-dimensional (quasi-2D, with finite thickness) collinear antiferromagnetic (AFM) system as a prototypical exemplary platform. To enable noncollinear magnetic order and engineer the NR Edelstein effect, we introduce a canted spin configuration with finite vector spin chirality $\bm{\kappa}=\hat{\bm{S}}_1\times\hat{\bm{S}}_2$, where $\hat{\bm S}_1$ and $\hat{\bm S}_2$ are the SAM unit vectors of the two magnetic sublattices in a unit cell \cite{Kipp2021ChiralHall,Wu2025ChiralSHG}. In order to explore the interplay between $\bm\kappa$ and Edelstein effect coefficients, we adopt a widely-studied quasi-2D honeycomb lattice and perform tight-binding model simulations. Applying the spin group-theoretical analysis, we enumerate the LEE and NLEE responses from both SAM and OAM accumulations (Fig. \ref{fig:schematic}). We find that all NR-LEE responses are symmetrically forbidden regardless of $\bm\kappa$. This is because the Zeeman quantum geometry generally vanishes in the NR limit. Hence, we further explore the NR-NLEE, which shows finite $\mathcal{T}$-odd SAM and $\mathcal{T}$-even OAM accumulation components. These NR-NLEE responses arise from the exchange field, hence is generally stronger than the SOC-induced NLEE components. Our results also show that vector spin chirality serves as an effective switching parameter to modulate the NLEE. In addition, we conduct \textit{ab initio} calculations for an exemplary realistic material, a quasi-2D $\rm MnSe$ monolayer \cite{MnSeAapro,MnSeSattar,Liu2023npjCM,MnSeQayyum,Wang2025MnSeValley}. We find that its SAM-contributed NR-NLEE could reach $\sim 10\,\mu_B\angstrom^2/\mathrm{V}^{2}$ with a small spin chirality, which could be experimentally detected under moderate electric field strength. Our results suggest that noncollinear spin textures, such as spin canting in antiferromagnets, provide a heavy-metal-free and field-tunable route toward charge-current-induced magnetization in 2D magnetic materials.

\section{Computational Methods}
\subsection{Density functional theory calculations}
The first-principles density functional theory (DFT) calculations are performed in the Vienna \textit{ab initio} simulation package (\textsc{vasp}) \cite{KresseFurth1996} using the projector augmented-wave method \cite{Bloechl1994,KresseJoubert1999} and the Perdew-Burke-Ernzerhof exchange-correlation functional \cite{PBE1996}. The correlations on the Mn $d$ orbital are treated within generalized gradient approximation with Hubbard $U$ corrections, namely, GGA$+U$ in the Dudarev scheme \cite{Dudarev1998}, with an effective $U=2.3~\rm eV$ according to previous studies on the same material \cite{MnSeSattar,MnSeQayyum}. Note that other $U$ values would not qualitatively change the main results. The quasi-2D monolayer is modeled with a hexagonal cell with a vacuum space of $20\,\angstrom$ perpendicular to the 2D plane to suppress interlayer interactions. Unless explicitly stated, SOC is completely turned off in the calculations. In order to evaluate the LEE and NLEE response functions, we construct maximally localized Wannier functions from the Mn $d$ and Se $p$ orbitals using the \textsc{wannier90} package \cite{MarzariVanderbilt1997,SouzaMarzari2001,Wannier90}. Finite spin chirality is introduced by applying a spin canting on the local magnetic moment of each $\rm Mn$. Such a spin chirality could occur naturally in realistic antiferromagnets through the Dzyaloshinskii--Moriya interaction \cite{Moriya1960,Wang21PhysRevLett.127.117202} or under an applied external magnetic field, and has recently been exploited to generate out-of-plane spin torques and nonlinear optical responses (e.g., bulk photovoltaic effects and second harmonic generation) in canted collinear antiferromagnets \cite{RuO2canted,tian2026observationmagneticallyswitchablequantum}.

\subsection{Edelstein effects}
We evaluate the charge-moment conversion using the electric field induced SAM and OAM accumulations, which are calculated according to the Kubo perturbative theory within the constant carrier relaxation time ($\tau$) approximation. Under an external electric field, the induced SAM accumulation can be expressed in terms of the LEE and NLEE as
\begin{equation}
    \begin{split}
        \delta s_a=&\alpha_{ab}E_b,\\
        \delta s_a=&\alpha_{abc}E_bE_c,
    \end{split}
\end{equation}
where $a$ labels the induced magnetization component direction, and $b$ (or $c$) refers to the electric field directions. For quasi-2D materials, we constrain the electric field direction in the $xy$ plane. Note that for the out-of-plane field, it is the displacement field $D_z$, not the electric field $E_z$, to serve as the natural variable \cite{Stengel2009NatPhys}. Thus, it is beyond the scope of the current study. We use different number of subscript indices to label the LEE and NLEE response tensors. The Einstein summation convention is adopted. Both LEE and NLEE susceptibility functions can be furthermore decomposed into $\mathcal{T}$-even and $\mathcal{T}$-odd parts. At the long relaxation time and clean limit, they are
\begin{equation}
    \begin{split}
        \tilde{\alpha}_{ab}^{\rm even}=&eV_{\rm u.c.}\int[d\bm k]\sum_{n}\mathrm{Re}(s_{nn}^av_{nn}^b)f_n',
    \end{split}
\end{equation}
and
\begin{equation}
    \begin{split}
        \alpha_{ab}^{\rm odd}=&\frac{eV_{\rm u.c.}}{\hbar}\int[d\bm k]\sum_{m,n}f_{nm}\mathrm{Im}(s_{nm}^av_{mn}^b)\frac{\omega_{nm}^2-\tau^{-2}}{[\omega_{nm}^2+\tau^{-2}]^2}.
    \end{split}
\end{equation}
Here, $f_{nm}=f_{n\bm k}-f_{m\bm k}$ and $\omega_{nm}=\omega_{n\bm k}-\omega_{m\bm k}$ is Fermi-Dirac occupation and eigenfrequency differences between band $n$ and $m$ at momentum $\bm k$, respectively. $f_n'=\partial f_n/\partial E$ refers to Fermi-surface contribution. $s_{nm}^a=\langle u_{n\bm k}|\hat{\sigma}^a|u_{m\bm k}\rangle\mu_B$ and $v_{mn}^b=\langle u_{m\bm k}|\hat{v}^b|u_{n\bm k}\rangle$ are the SAM and velocity operator matrix, respectively. The integral $\int[d\bm k]=\int\frac{d^D\bm k}{(2\pi)^D}$ is performed in the $D$-dimensional first BZ. $V_{\rm u.c.}$ is the total $D$-dimensional volume of the unit cell. Hence, the components evaluate the total magnetic moments accumulated in a single unit cell. The $\mathcal{T}$-even term essentially denotes the accumulation rate that saturates at carrier lifetime $\tau$. Hence, we denote it as $\tilde{\alpha}^{\rm even}$, while the $\mathcal{T}$-odd term is $\tau$-independent and denoted as $\alpha^{\rm odd}$. They differ by a unit of time. Such a denotation is also used for the NLEE components. Microscopically, one can show that they both scale with Zeeman quantum geometry tensor $\mathfrak{Q}_{nm}^{ab}(\bm k)=-\langle u_{n\bm k}|i\partial_{k_a}|u_{m\bm k}\rangle\langle u_{m\bm k}|\sigma_b|u_{n\bm k}\rangle$ in the mixed momentum-spin space \cite{Xiang2025PRL,Cao2025PRB,Chakraborti2025arXiv}.

The $\mathcal{T}$-even part of NLEE arises from the intrinsic ASP-dipole in the form of \cite{Xiao2023PRL130},
\begin{equation}
    \tilde{\alpha}^{\rm even}_{abc}=\frac{e^2V_{\rm u.c.}}{\hbar^2}\int[d\bm k]\sum_nf_n\partial_{k_c}\Upsilon^{ab}_n,
\end{equation}
where $\Upsilon_{n}^{ab}(\bm{k})= 2{\rm Im}\sum_{m\neq n}\frac{s^a_{nm}v^b_{mn}}{\omega_{nm}^2}$ is the ASP. The $\mathcal{T}$-odd part is \cite{Gao2014,Xiao2022PRL129},
\begin{equation}
    \begin{split}
        \alpha^{\rm odd}_{abc}=&-\frac{e^2V_{\rm u.c.}}{2\hbar}\partial_{h_a}\int[d\bm k]\sum_nG^{bc}_nf_n \\
         &-\frac{e^2V_{\rm u.c.}}{\hbar}\int[d\bm{k}]\sum_n(s^a_{nn} G^{bc}_n+v^b_{nn} \mathcal{G}^{ac}_n) f_n' ,
    \end{split}
\end{equation}
where $G^{bc}_n= 2{\rm Re}\sum_{m\neq n}\frac{v^b_{nm}v^c_{mn}}{\omega_{nm}^3}$ is the conventional $\bm{k}$-space BCP, and $\mathcal{G}^{ac}_n=-2{\rm Re}\sum_{m\neq n}\frac{s^a_{nm}v^c_{mn}}{\omega_{nm}^3}$ the Zeeman-like BCP in the mixed spin-$\bm{k}$ space. Considering the interband position matrix element $r_{nm}^a=i\langle u_{n\bm k}|\partial_{k_a}|u_{m\bm k}\rangle$, one can rewrite the ASP term as $\Upsilon_{n}^{ab}(\bm{k})= 2{\rm Re}\sum_{m\neq n}
\mathfrak{Q}_{nm}^{ab}(\bm k)/\omega_{mn}$. Hence, the $\mathcal{T}$-even NLEE is a Zeeman-ASP dipole response; the Zeeman-like BCP as $\mathcal{G}_{ac}^n(\bm k)=-2{\rm Im}\sum_{m\neq n}\mathfrak{Q}_{mn}^{ca}(\bm k)/\omega_{nm}^2$. This indicates that the $\mathcal{T}$-odd NLEE is essentially a Zeeman-like BCP. In general, both of them can be activated by the exchange field in the NR limit.

In addition to the SAM contributions to magnetic moment, we also estimate the OAM contributions. Note that the nonequilibrium behaviors of OAM can be as significant as their SAM counterparts, as suggested by recent model and first-principles studies \cite{LeivaMontecinos2023,Dong2024,Go20PRResearch,Mu2021npjCM,Xu21PRB,Yang2025PRB}, as well as experimental measurements \cite{Ding2022OREE,Jo24npjSpin}. In our current work, this is done by replacing the spin operator $\hat{\sigma}$ by OAM operator $\hat{L}$, and the corresponding LEE and NLEE tensors are denoted as $\tilde{\beta}^{\rm even}$ and $\beta^{\rm odd}$. According to the modern theory of OAM, one has interband orbital moment operator elements $\langle u_{n\bm k}|\hat{L}^a|u_{m\bm k}\rangle=\epsilon_{abc}\frac{e\hbar}{2g_L\mu_B}\sum_{l\neq m,n}[v_{nl}^br_{lm}^c+(v_{nn}^b+v_{mm}^b)r_{nm}^c]$, where $\epsilon_{abc}$ the Levi-Civita symbol. This includes not only the intra-atomic orbital contribution, but also the inter-atomic cycloid motions. Note that for the quasi-2D systems the out-of-plane $z$ direction is not periodic. The in-plane OAM then arises from the inter-atomic layer tunneling. Unlike the position operator along the periodic boundary direction that is encoded using the derivative structure of the crystal wavefunction, i.e., Berry connection, the out-of-plane non-periodic position operator can be naturally defined using the Wannier center position operator $r_{nm}^z=\mathcal{Z}_{nm}$. We follow the previously proposed strategy \cite{Ghorai2025PRL,Biswas2025PRB} to compute the $z$-component velocity matrix using
\begin{equation}
    v_{nm}^z=i\omega_{nm}\mathcal{Z}_{nm}.
\end{equation}
Hence, one can estimate the in-plane OAM for quasi-2D systems with nonzero thickness. We note that the derivative of quantum metric contributions to orbital NLEE effects, especially the $\partial_{k_z}$-related terms \cite{Qiang2026PRL,Qian2026PRB}, are not included here, which do not affect the symmetry analysis and the main conclusions.

\section{Results and Discussions}
\subsection{Tight-binding model and magnetic structure}
\begin{figure}[t]
\centering
\includegraphics[width=0.95\columnwidth]{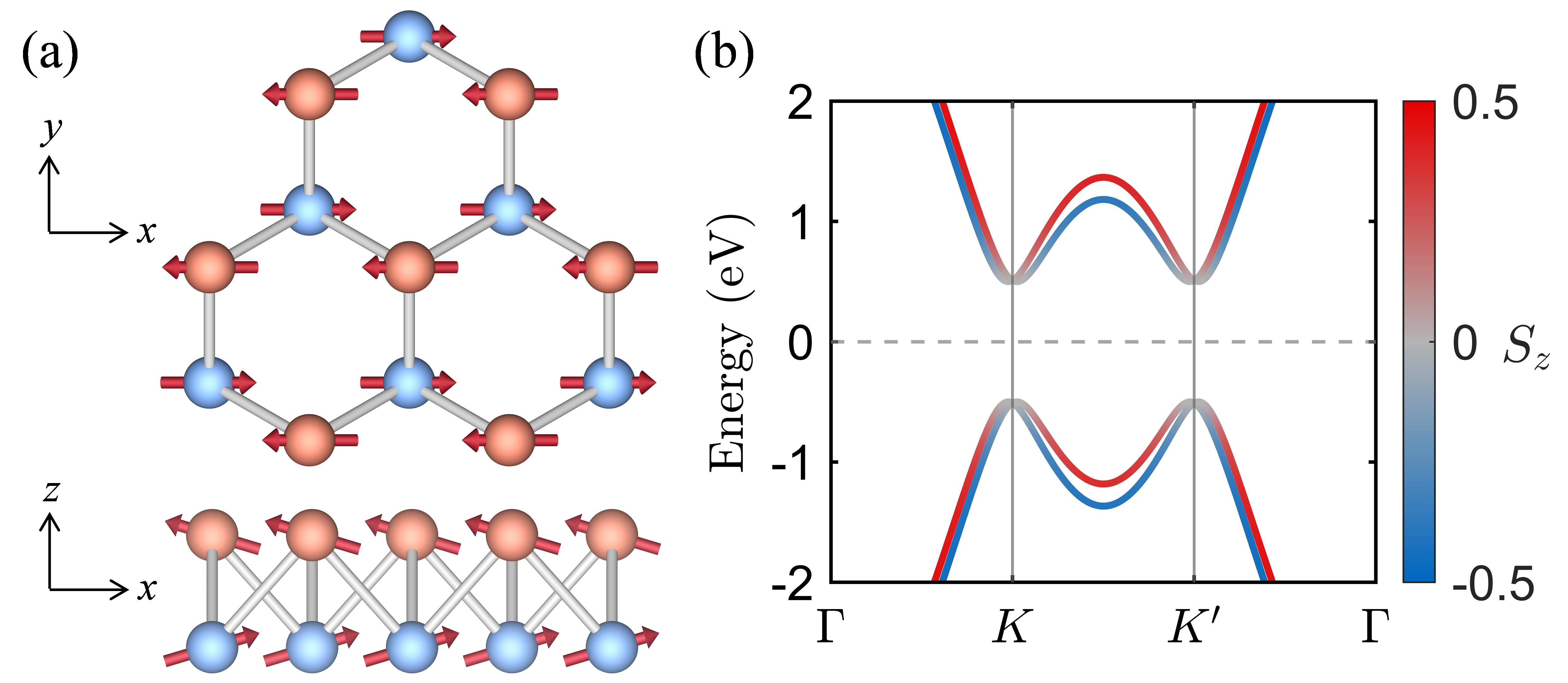}
\caption{(a) Top and side views of quasi-2D honeycomb model with vector spin chirality $\hat{\bm\kappa}$ along $\hat{\bm y}$. The red arrows indicate local spin polarization on each sublattice. (b) Tight-binding band dispersion along the high-symmetric $k$ path. The expectation value of $z$ component SAM  (along the $\hat{\bm m}$ direction, in unit of $\hbar$)  is encoded by color from blue to red. We take typical parameters to conduct the simulations, $V_{pp\pi}=1\,\rm eV$, $V_{pp\sigma}=1.5\,\rm eV$, and $S=0.5\,\rm eV$. The lattice constant and the quasi-2D system thickness are taken to be $2.46\,\angstrom$ and $1.0\,\angstrom$.}
\label{fig:TB}
\end{figure}

We start our analysis and simulation from a tight binding model of a quasi-2D buckled AFM honeycomb lattice [Fig. \ref{fig:TB}(a)]. Each unit cell contains two sites $A$ and $B$ with alternating height along $z$. Using the basis set of $[|p_z^A\uparrow\rangle,|p_z^B\uparrow\rangle,|p_z^A\downarrow\rangle,|p_z^B\downarrow\rangle]^{\rm T}$, the Hamiltonian is
\begin{equation}
H(\bm{k}) =t\sum_{\langle i,j\rangle}c_i^{\dagger}c_j\otimes\sigma_0 + \sum_i \hat{\bm{S}}_i\cdot\bm{\sigma},
\label{eq:Ham}
\end{equation}
where $t=\cos^2\theta_{ij}V_{pp\pi}+(1-\cos^2\theta_{ij})V_{pp\sigma}$ is Slater-Koster type hopping integral between nearest neighbor site pairs $i$ and $j$ ($\theta_{ij}$ is the elevation angle between them, with respect to the $xy$ plane). $\sigma_0$ and $\bm{\sigma}$ are the identity and Pauli matrices in spin space, and $\hat{\bm{S}}_i$ refers to the staggered local exchange field on sublattice $i$. In principle, one can denote the N\'eel direction as $\hat{\bm n}$ and apply a spin canting along its perpendicular direction $\hat{\bm m}$. Then, we achieve a noncollinear (coplanar) spin configuration characterized by the vector spin chirality $\bm \kappa$ that is normal to N\'eel vector and canted spin directions. Since SOC is absent, the spin rotation symmetry is preserved. Therefore, in principle, one can \textit{define} a spin-space coordinate frame based upon $\hat{\bm n}$-$\hat{\bm m}$-$\hat{\bm\kappa}$, where 
\begin{equation}
    \begin{split}
        \hat{\bm n}=&\frac{\hat{\bm S}_A-\hat{\bm S}_B}{|\hat{\bm S}_A-\hat{\bm S}_B|},\\
        \hat{\bm m}=&\frac{\hat{\bm S}_A+\hat{\bm S}_B}{|\hat{\bm S}_A+\hat{\bm S}_B|},\\
        \hat{\bm\kappa}=&\hat{\bm n}\times\hat{\bm m}.
    \end{split}
\end{equation}
Throughout this work, without loss of generality, we align the local exchange field along the $\hat{\bm n}=\hat{\bm x}$ direction, i.e., the collinear spin configuration is given by $\hat{\bm{S}}_{A,B}=(\pm \hat{S},0,0)$. We then introduce a spin canting along $\hat{\bm m}=\hat{\bm z}$, resulting in $\hat{\bm{S}}_A = S(-\cos\theta,0,\sin\theta)$, and $\hat{\bm{S}}_B = S(\cos\theta,0,\sin\theta)$. Accordingly, the vector spin chirality is along $\hat{\bm y}$, written as
\begin{equation}
    \bm{\kappa} =\frac{1}{S^2}(\hat{\bm{S}}_A\times\hat{\bm{S}}_B) = \sin(2\theta)\hat{\bm{y}} ,
\label{eq:kappa}
\end{equation}
which serves as the order parameter in this work.

\subsection{Spin-group symmetry and selection rules}
Before presenting simulation results, we first perform a spin group-theoretical analysis of
the band resolved spin texture and (N)LEE responses for both spin and orbital degrees of freedom. Each spin group symmetry operator can be written as $[R_s||R_l]$ with $R_s$ and $R_l$ being the individual operators on the spin and lattice spaces, respectively. When the chiral spin is not present ($\kappa=0$), the collinear magnetic pattern is described by the direct product between its spin only group $\mathbb{G}_{\rm so}=\mathrm{SO}(2)\rtimes Z_2^{K}$ and nontrivial spin group $\mathbb{G}_{\rm nsp}$, where $Z_2^{K}=\{E,\bar{C}_{2\perp}\}$ is generated by a $180^{\circ}$ rotation vertical to the N\'eel direction ($\hat{\bm n}$). In the buckled honeycomb lattice, its crystalline point group is $\bar{3}m$. One finds that the collinear $\mathbb{G}_{\rm nsp}$ is $^{\bar{1}}\bar{3}^1m$ that contains $[E||C_{3z}]$, $[E||M_x]$, and $[C_{2z}||\bar{E}]$ here.

\begin{table*}[t]
\caption{Symmetry constraints on the LEE and NLEE components of the spin-canted state with $\bm{\kappa}\parallel\hat{\bm{y}}$ magnetic configuration. We use \checkmark~symbol to denote symmetry-allowed components without SOC (NR-limit, allowed by spin group), $\diamond$ for the ones are forbidden without SOC but allowed with SOC (forbidden in spin group but allowed in magnetic group, i.e., $2'$ in this situation), and \xmark~for the components completely forbidden even with SOC (from magnetic group).}
\label{tab:selrules}
\begin{ruledtabular}
\begin{tabular}{lcccccc}
\multicolumn{7}{c}{ LEE, $\delta M_a=\alpha_{ab}E_b$ or $\delta M_a=\beta_{ab}E_b$} \\
\colrule
 & $xx$ & $xy$ & $yx$ & $yy$ & $zx$ & $zy$ \\
$\alpha^{\rm odd}$  & \xmark   & $\diamond$ & $\diamond$ & \xmark   & $\diamond$ & \xmark   \\
$\tilde{\alpha}^{\rm even}$ & $\diamond$ & \xmark   & \xmark   & $\diamond$ & \xmark   & $\diamond$ \\
$\beta^{\rm odd}$    & \xmark   & $\diamond$ & $\diamond$ & \xmark   & $\diamond$ & \xmark   \\
$\tilde{\beta}^{\rm even}$ & $\diamond$ & \xmark   & \xmark   & $\diamond$ & \xmark   & $\diamond$ \\
\end{tabular}
\end{ruledtabular}

\vspace{6pt}

\begin{ruledtabular}
\begin{tabular}{lccccccccc}
\multicolumn{10}{c}{ NLEE, $\delta M_a=\alpha_{abc}E_bE_c$ or $\delta M_a=\beta_{abc}E_bE_c$}\\
\colrule
 & $xxx$ & $xxy$ & $xyy$ & $yxx$ & $yxy$ & $yyy$ & $zxx$ & $zxy$ & $zyy$ \\
$\alpha^{\rm odd}$  & \xmark   & $\diamond$ & \xmark   & $\diamond$ & \xmark   & $\diamond$ & $\checkmark$ & \xmark   & $\checkmark$ \\
$\tilde{\alpha}^{\rm even}$  & $\diamond$ & \xmark   & $\diamond$ & \xmark   & $\diamond$ & \xmark   & \xmark     & $\diamond$ & \xmark     \\
$\beta^{\rm odd}$    & \xmark   & $\diamond$ & \xmark   & $\diamond$ & \xmark   & $\diamond$ & $\diamond$   & \xmark   & $\diamond$   \\
$\tilde{\beta}^{\rm even}$   & $\checkmark$ & \xmark & $\checkmark$ & \xmark & $\checkmark$ & \xmark   & \xmark     & $\diamond$ & \xmark     \\
\end{tabular}
\end{ruledtabular}
\end{table*}

One can adopt the Neumann's principle to analyze the symmetry constraints on both spin LEE and NLEE components,
\begin{equation}
    \begin{aligned}
    \alpha_{ab}  &= \eta_s R_{s}^{aa'}R_{l}^{bb'}\alpha_{a'b'}\\
    \alpha_{abc} &= \eta_s R_{s}^{aa'}R_{l}^{bb'}R_{l}^{cc'}\alpha_{a'b'c'},
    \end{aligned}
    \label{eq:spininvariance}
\end{equation}
where $\eta_s$ is determined by the effective time-reversal symmetry $\mathcal{T}$ in the system. When $R_s$ is proper rotation ($\det(R_s)=1$), the $[R_s||R_l]$ is $\mathcal{T}$-even, and $\eta_s=1$. However, if $\det(R_s)=-1$, $[R_s||R_l]$ contains an effective $\mathcal{T}$, then $\eta_s=\pm 1$ is taken for $\alpha^{\rm odd}$ and $\tilde{\alpha}^{\rm even}$ components, respectively. This is because in the spin group framework, the spin vector is treated as a polar vector under $R_s$. In other words, for $\alpha^{\rm odd}$ one always has $\eta_s=1$, while for $\tilde{\alpha}_{\rm even}$, $\eta_s=\det(R_s)$. The orbital LEE and NLEE components ($\beta_{ab}$ and $\beta_{abc}$) can be determined as well. By contrast, it is only the $R_l$ part that controls them, as spin and lattice are completely decoupled. This yields 
\begin{equation}\label{eq:orbinvariance}
    \begin{split}
        \beta_{ab}=&\eta_l\det(R_l)R_{l}^{aa'}R_{l}^{bb'}\beta_{a'b'}\\
        \beta_{abc}=&\eta_l\det(R_l)R_{l}^{aa'}R_{l}^{bb'}R_{l}^{cc'}\beta_{a'b'c'},
    \end{split}
\end{equation}
where $\det({R_l})$ encodes the proper or improper rotation of orbital angular momentum. $\eta_l=1$ for $\tilde{\beta}^{\rm even}$, and $\eta_l=\det{(R_s)}$ for $\beta^{\rm odd}$.

According to Eqs. (\ref{eq:spininvariance}) and (\ref{eq:orbinvariance}), we find that all the LEE and spin NLEE are symmetrically forbidden in the collinear AFM configuration here. Hence, we introduce finite vector spin chirality $\kappa_y\neq 0$ to reduce the symmetry. In this situation, the spin only group becomes $\mathbb{G}_{\rm so}=^{m_y}1$ and we have $\mathbb{G}_{\rm nsp}$ to be $^{2_z}\bar{3}^1m$. Once again, we find that LEE components are always forbidden under spin group constraints. Interestingly, there are two symmetry-allowed spin NLEE ($\alpha_{zxx}^{\rm odd}=\alpha_{zyy}^{\rm odd}$) and three orbital NLEE components ($\tilde{\beta}_{xxx}^{\rm even}=-\tilde{\beta}_{xyy}^{\rm even}=-\tilde{\beta}_{yxy}^{\rm even}$). Both of them only contain one independent element. All spin $\tilde{\alpha}^{\rm even}$ and orbital $\beta^{\rm odd}$ channels remain forbidden in the SOC-free limit. The $\mathcal{T}$-even and $\mathcal{T}$-odd separation between the SAM and OAM degrees of freedom yields a clear strategy to distinguish them through flipping N\'eel vector $\hat{\bm n}$ or spin canting direction $\hat{\bm m}$. We tabulate the symmetry constraints of all these components under in-plane electric field in Table \ref{tab:selrules}. It should be emphasized that this table is based upon the N\'eel vector $\hat{\bm n}$ along $\hat{\bm x}$, spin canting $\hat{\bm m}$ along $\hat{\bm z}$, and hence vector spin chirality $\hat{\bm\kappa}$ along $\hat{\bm y}$. Under spin rotational symmetry, the LEE and NLEE components of the spin part can be denoted in the $\hat{\bm n}-\hat{\bm m}-\hat{\bm\kappa}$ frame. For instance, the spin-$x$ in $\alpha_{xyy}$ essentially indicates spin accumulation along the N\'eel vector direction (or in general $\alpha_{\hat{\bm n}yy})$, and $\alpha_{zxx}$ means spin along canted magnetic moment direction $\hat{\bm m}$. For OAM accumulation ($\beta$'s), since it is well controlled by lattice symmetry, all indices are in line with the real lattice Cartesian coordinate framework. When SOC is included, the spin and lattice coordinates are interlocked and a universal coordinate framework should be adopted.

\subsection{Chirality engineering electronic structure and Edelstein effects}
Armed with the spin symmetry analysis, we next examine the electronic band dispersion and spin texture of the canted configuration, as plotted in Fig. \ref{fig:TB}(b). The magnetic exchange field leads to a finite direct bandgap of $1.0\,\rm eV$. In the collinear magnetic pattern, both valence and conduction bands exhibit Kramers degeneracy, which can be lifted upon introducing finite vector spin chirality. Here, one sees that $\langle s_z\rangle(\bm k)=\langle s_z\rangle(-\bm k)$, constrained by $[\bar{C}_{2x}||\bar{E}\mathcal{T}]$. The in-plane spin components are forbidden in the model Eq.  (\ref{eq:Ham}). The band-resolved OAM components, due to their time-reversal odd nature, completely vanish under $[\bar{C}_{2x}||\bar{E}\mathcal{T}]$. Hence, even though the equilibrium OAM is quenched under spin group constraints, the non-equilibrium OAM accumulation could arise (see below).

\begin{figure}[t]
\centering
\includegraphics[width=1.0\columnwidth]{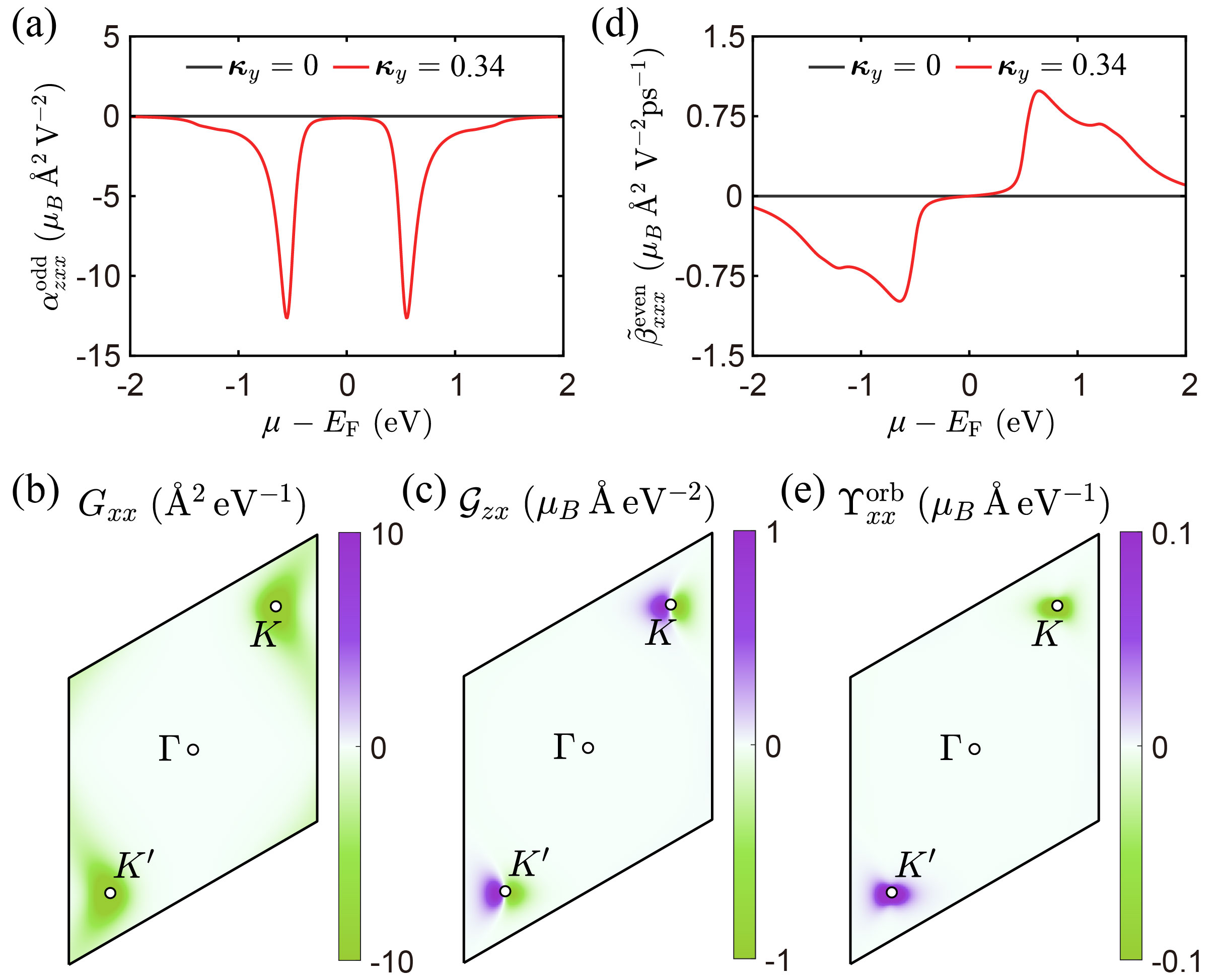}
\caption{(a) $\mathcal{T}$-odd spin-NLEE component $\alpha^{\rm odd}_{zxx}$ variation with respect to chemical potential $\mu-E_{\rm F}$ in the SOC-free limit, for the collinear ($\kappa_y=0$, black) and canted ($\kappa_y=0.34$, red) configurations. (b) and (c) plot the momentum distribution of BCP ($G^{xx}(\bm k)$) and Zeeman BCP ($\mathcal{G}^{xx}(\bm k)$) of the valence bands. (d) The $\mathcal{T}$-even orbital-NLEE component $\tilde{\beta}_{xxx}^{\rm even}$ under different chemical potential. (e) The $\bm k$-resolved anomalous orbital polarizability $\Upsilon^{xx}(\bm k)$ distribution, whose dipole moment controls the $\tilde{\beta}_{xxx}^{\rm even}$.}
\label{fig:wosoc-mu}
\end{figure}

Next, we plot the dominant spin $\mathcal{T}$-odd NLEE component $\alpha^{\rm odd}_{zxx}=\alpha^{\rm odd}_{zyy}$ as a function of chemical potential $\mu$ in Fig.~\ref{fig:wosoc-mu}. In the collinear state, the responses vanish identically across the entire energy window, confirming their symmetry prohibition. It is clear that spin canting serves as an effective knob to induce spin-NLEE. One sees that when $\kappa=0.34$ (canting angle of $10^{\circ}$), $\alpha_{zxx}^{\rm odd}$ reaches $-13\,\mu_B\angstrom^2/\mathrm{V}^{2}$ at the band edge ($\mu-E_F=\pm 0.55\,\rm eV$) [Fig.~\ref{fig:wosoc-mu}(a)]. The valence and conduction band contributions are the same due to the particle-hole symmetry in the tight binding model.

To identify the microscopic origin of the response, we plot the momentum-resolved BCP $G^{xx}_{n}$ and Zeeman-like BCP $\mathcal{G}_n^{xx}$  of the valence band in Figs.~\ref{fig:wosoc-mu}(b) and \ref{fig:wosoc-mu}(c), respectively. One sees that pronounced peaks appear in the vicinity of the $K$ and $K'$ valleys, reflecting the strong enhancement of these band-geometric quantities, which scale as $\omega_{nm}^{-3}$ and become significantly amplified near band-degenerate points. Consequently, the spin-NLEE reaches its maximum value near the band edge.

The orbital NLEE $\tilde{\beta}^{\rm even}_{xxx}=-\tilde{\beta}_{xyy}^{\rm even}=-\tilde{\beta}_{yxy}^{\rm even}$ as a function of chemical potential is plotted in Fig.~\ref{fig:wosoc-mu}(d), reaching a value of $0.8~\mu_B\angstrom^2/(\rm V^2\cdot ps)$ near the band edge in the canting structure. On the contrary to spin-NLEE, one sees that the response function is positive and negative when $\mu$ is above and below the Fermi energy, indicating opposite behaviors for electron and hole carriers. The momentum distribution of valence band anomalous orbital polarizability shows contrasting valley distributions [Fig. \ref{fig:wosoc-mu}(e)].

\begin{figure}[t]
\centering
\includegraphics[width=0.8\columnwidth]{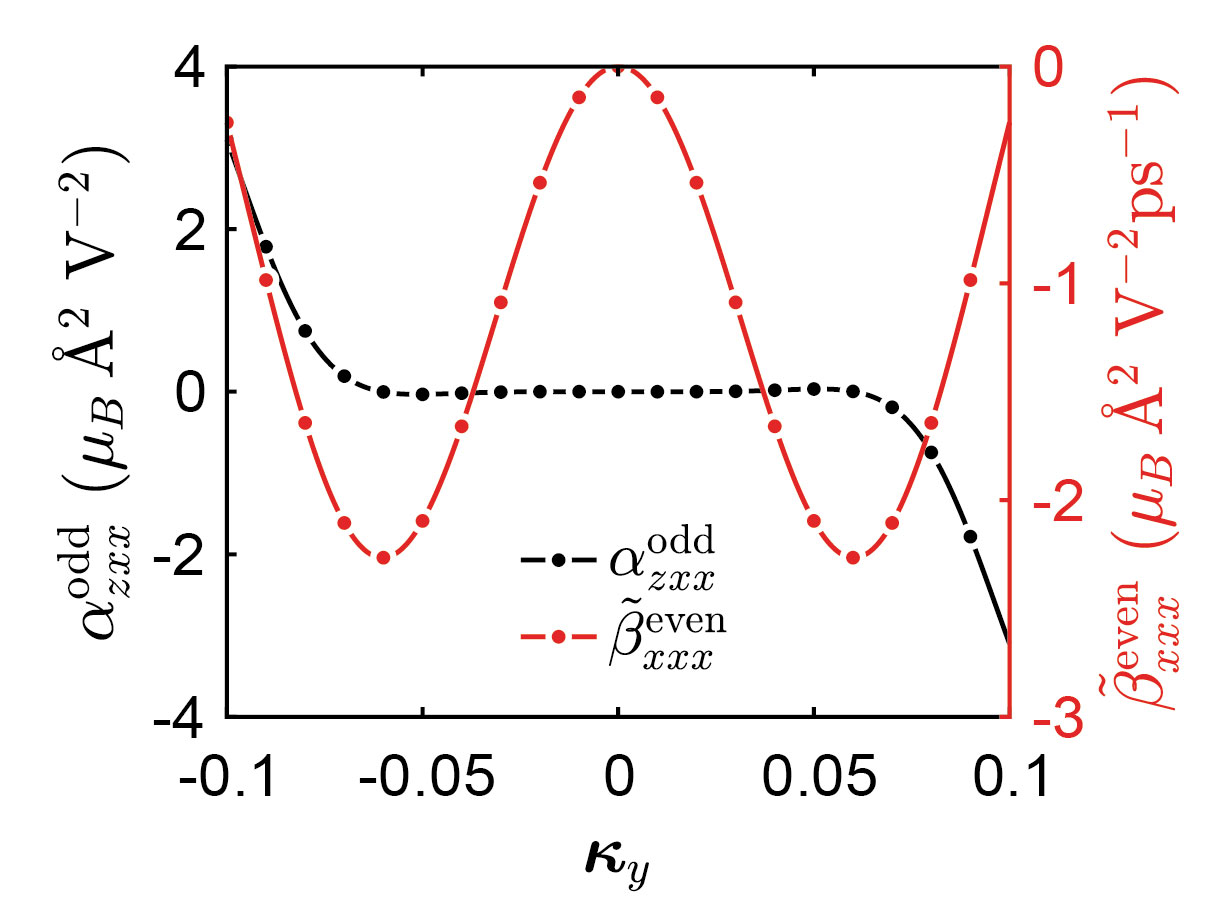}
\caption{Vector spin chirality controlled $\mathcal{T}$-odd  spin-NLEE component $\alpha^{\rm odd}_{zxx}$ (black, left axis) and $\mathcal{T}$-even orbital-NLEE component $\beta^{\rm even}_{yxy}$ (red, right axis). Both of their values are taken from a typical chemical potential $\mu=-0.8\,\rm eV$. They exhibit odd and even functions with respect to $\kappa$.}
\label{fig:wosoc-kappa}
\end{figure}

We argue that one can use vector spin chirality to effectively tune the NLEE responses. When the spin canting direction is flipped (with the equilibrium N\'eel vector fixed), it is equivalent to apply a $[\bar{C}_{2z}||\mathcal{T}]$ operator onto the system. In such a circumstance, it is clear that the nonzero spin-NLEE can be reversed as $[\bar{C}_{2z}||\mathcal{T}]\alpha_{zxx}^{\rm odd}=-\alpha_{zxx}^{\rm odd}$ [Eq. (\ref{eq:spininvariance})], which is an odd function with $\bm \kappa$. On the contrary, since $[\bar{C}_{2z}||\mathcal{T}]\tilde{\beta}_{xyy}^{\rm even}=\tilde{\beta}_{xyy}^{\rm even}$ [Eq. (\ref{eq:orbinvariance})], the nonzero orbital-NLEE remains unchanged upon reversing $\bm\kappa$. In order to illustrate such a different behavior, we continuously adjust $\kappa_y$ in the model, and track the spin and orbital NLEE responses, as plotted in Fig. \ref{fig:wosoc-kappa}. Our simulation results are well consistent with the symmetry argument. Note that in realistic experiments it is usually challenging to distinguish SAM and OAM contributions, as they both contribute to magnetic responses during nonequilibrium processes. While the spin accumulation can be detected by Kerr rotation or nonlocal spin-valve and spin Hall geometries, orbital accumulation is usually inferred through orbital-to-spin conversion or orbital-torque measurements. These strategies are still not direct measurements of OAM responses. Recent experiments infer orbital accumulation and transport indirectly, such as by using weak-SOC orbital-current sources combined with complementary analysis, and such separation generally remains model dependent \cite{Carra1993PRL,Boeglin2010Nature,Ding2020PRL,Lee2021NatCommun,Go2021EPL}. Based on our above analysis, the distinct responses of SAM and OAM accumulations under reversal of vector spin chirality provide an additional symmetry-based criterion for experimentally distinguishing these two contributions in noncollinear magnetic systems.

\subsection{First-principles results for $\rm MnSe$ monolayer}
\begin{figure}[t]
\centering
\includegraphics[width=1.0\columnwidth]{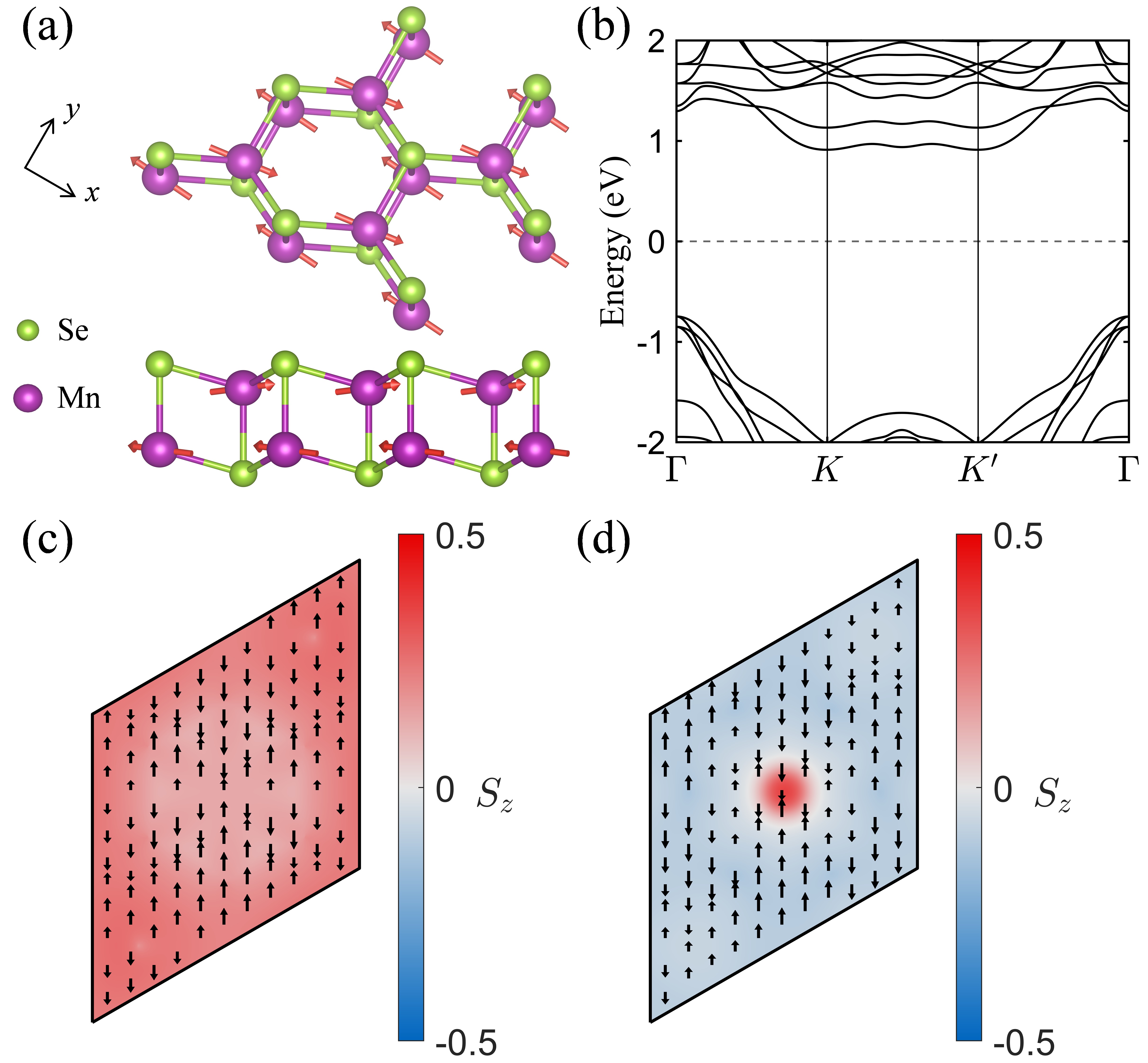}
\caption{(a) Top and side view of $\rm MnSe$ monolayer with the local magnetic moment on $\rm Mn$ indicated by the arrows. (b) Band structure along $\Gamma$--$K$--$K'$--$\Gamma$, with the color scale encoding the out-of-plane spin polarization $\langle s_z\rangle$ of each state. Spin texture of (c) the top valence band and (d) bottom conduction band. The arrows indicate in-plane spin texture, while the out-of-plane spin texture is encoded by the background color.}
\label{fig:dft-bands}
\end{figure}

In order to illustrate and quantify the proposed effects in realistic materials, we perform first-principles calculations on the electronic structure and magnetic properties of a representative $\rm MnSe$ monolayer [Fig. \ref{fig:dft-bands}(a)]. It is a quasi-2D system with two magnetic $\rm Mn$ atoms in one unit cell, forming a buckled honeycomb lattice. The ground state shows an in-plane collinear AFM order \cite{MnSeAapro,MnSeSattar,Liu2023npjCM}. According to our calculations, the interplane and intraplane $\rm Mn$-$\rm Se$ bond lengths are relaxed to be $2.58$ and $2.60~\angstrom$, and its thickness is $3.32~\angstrom$. Each $\rm Mn$ atom carries about $4.38~\mu_B$ within its Wigner-Seitz cell. 
Given the collinear AFM configuration, the spin canting can be induced by applying an external magnetic field or by Dzyaloshinskii-Moriya interaction with the help of substrate or interface. Note that such a spin canting has been realized in various 2D antiferromagnets, such as $\rm MnPSe_3$ monolayer \cite{Zhang2024MnPSe3,Liu2024MnPSe3}, $\rm CrSBr$ bilayer \cite{Wu2025ChiralSHG,Wang2023CrSBr}, exfoliated $\rm FePS_3$ \cite{Feringa2022FePS3}, and few-layer $\rm CrPS_4$ \cite{Shi2024CrPS4}. Here, we introduce spin canting from the collinear AFM structure with N\'eel vector aligned along $\hat{\bm x}$ and track the total energy with respect to vector spin chirality. We find that the energy cost is $16.2~\rm meV$ per unit cell for a small canting angle of $10^{\circ}$ (or $\kappa=\pm 0.34$). Such a small energy value is within the experimentally realizable range.

Figure~\ref{fig:dft-bands}(b) shows the calculated band dispersion along the high symmetric $k$-path. It shows a semiconducting feature with the bandgap calculated to be $E_g\simeq1.8$~eV, in good agreement with previous calculations~\cite{MnSeSattar,MnSeQayyum}. Without considering SOC effect, the oriented spin space group is $P^{2_{001}}\bar{3}^{1}m^{m_{120}}1$ and its magnetic space group is $C2^{\prime}$. Consistent with the tight binding model, one sees that finite vector spin chirality lifts the Kramers degeneracy and exhibits finite spin expectation for each nondegenerate state. Here, $[\bar{C}_{2x}||\bar{E}\mathcal{T}]$ enforces $\langle s_x\rangle(\bm k)=0$, and the $[\bar{C}_{2y}||\mathcal{T}]$ protects odd and even parity of $\langle s_y\rangle(\bm k)=-\langle s_y\rangle(-\bm k)$ and $\langle s_z\rangle(\bm k)=\langle s_z\rangle(-\bm k)$ [Figs.  \ref{fig:dft-bands}(c) and \ref{fig:dft-bands}(d)]. The calculated band-resolved OAM components vanish due to the $\mathcal{T}$-odd nature of OAM.

\begin{figure}[t]
\centering
\includegraphics[width=1.0\columnwidth]{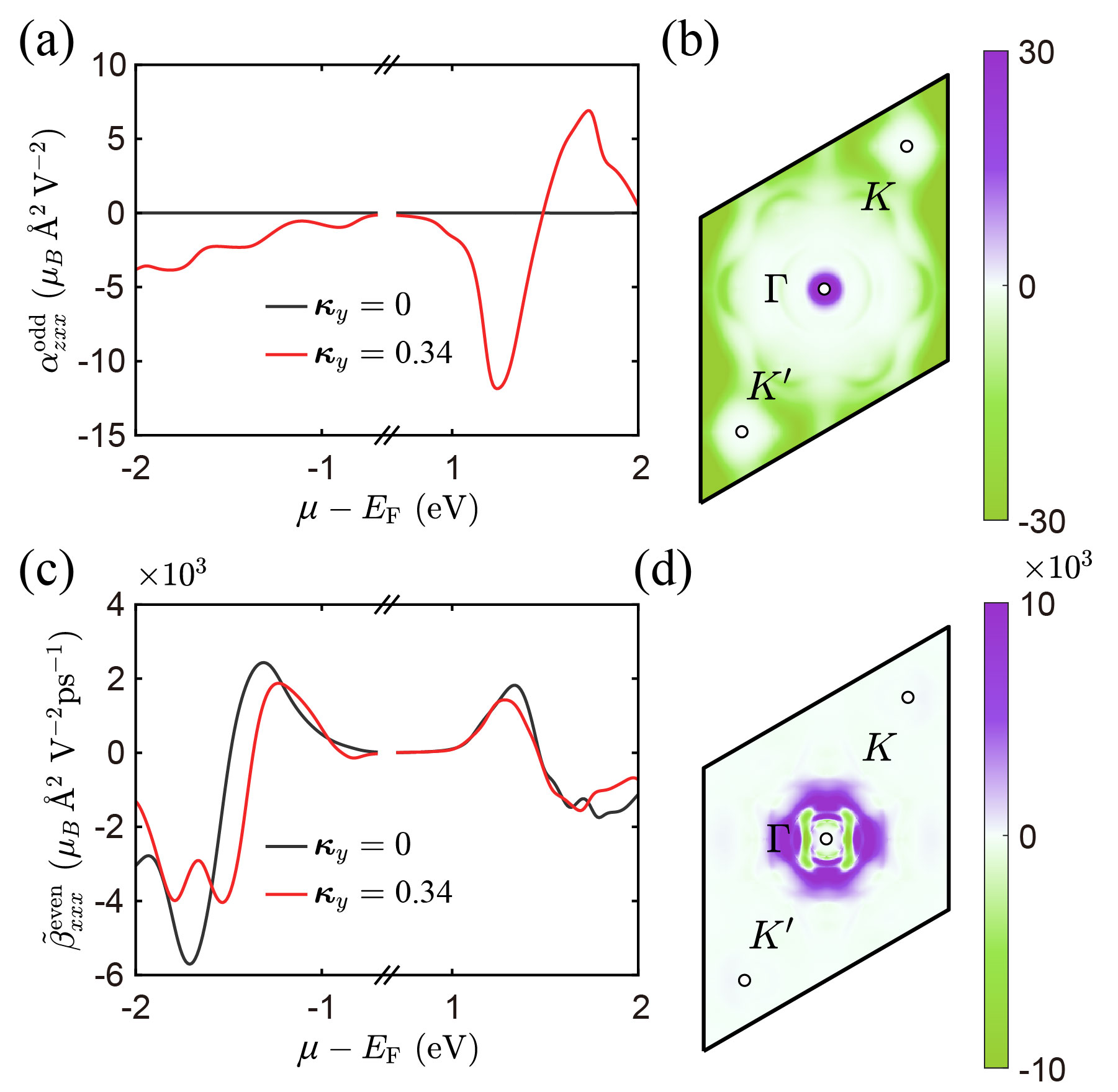}
\caption{(a) Calculated $\alpha^{\rm odd}_{zxx}$ of monolayer MnSe with respect to chemical potential, for the collinear ($\kappa_y=0$, black) and spin canting state ($\kappa_y=0.34$, red). (b) Momentum-distribution of $\alpha_{zxx}^{\rm odd}$ near the conduction band edge. (c) Chemical potential modulated $\tilde{\beta}_{xxx}^{\rm even}$ and (d) the $\bm k$-resolved contributions of the spin canting configuration near valence band edge.}
\label{fig:dft-response}
\end{figure}

We then calculate the NR Edelstein effects. The results are well consistent with previous symmetry analysis. In Fig. \ref{fig:dft-response}(a) we plot $\alpha_{zxx}^{\rm odd}$ in two typical $\kappa_y$ values, i.e., collinear pattern with $\kappa=0$ and spin canting state of $\kappa_y=0.34$. In the collinear state, $\alpha_{zxx}^{\rm odd}$ is identically zero. Under spin canting, it could reach $\simeq -12~\mu_B\angstrom^2/\mathrm{V}^{2}$ per unit cell at conduction band edge. This magnitude indicates that under a moderate electric field of $100\,\rm kV/cm$, the induced spin accumulation can reach $-1.2\times10^{-5}\,\mu_B$, well within the experimental detection capability. In order to see its BZ distribution, we calculate the integrand in Fig. \ref{fig:dft-response}(b). It shows that the dominant contribution comes near the BZ boundary, while compensated contributions appear around $\Gamma$.

As for the OAM accumulation, we plot $\tilde{\beta}_{xxx}^{\rm even}$ in Fig. \ref{fig:dft-response}(c). Unlike the spin accumulation, the orbital NLEE component remains finite even in the collinear limit ($\kappa = 0$). Upon introducing spin canting ($\kappa = 0.34$), this orbital response is strongly modified and exhibits a negative peak of $-4\times10^{3}~\mu_B\angstrom^2/(\mathrm{V}^{2}\cdot\mathrm{ps})$ near valence band edge. If we choose a typical carrier lifetime of $\tau$ to be $0.2\,\rm ps$, the peak of orbital accumulation is one order of magnitude larger than the maximum value of spin NLEE component for the same spin canting configuration with $\kappa = 0.34$. This clearly implies that the nonequilibrium OAM responses could become dominant over SAM responses, which has been largely overlooked previously. We plot its BZ distribution near the valence band edge, [Fig. \ref{fig:dft-response}(d)], which are mainly localized near the BZ center, consistent with its band dispersion feature.

As a potential experimental detection scheme, we note that such an electrically-induced
nonequilibrium magnetization can be probed through magneto-optical Kerr spectroscopy, anisotropic magnetoresistance, or nonlocal spin-valve detection~\cite{Johansson2024,Xiao2022PRL129,Xiao2023PRL130}. Since the $\mathcal{T}$-even and $\mathcal{T}$-odd contributions have different $\tau$ dependence (the former saturates at $\tau$ while the latter is $\tau$-independent), they can also be distinguished by varying the sample quality and/or environmental temperature. As discussed previously, one can also reverse vector spin chirality $\bm\kappa$ and track their odd/even feature to further disentangle their contributions.

\subsection{SOC effect}
Before concluding, we introduce the SOC effect and compare its impact on the NLEE response tensors. When the spin and lattice spaces are locked with finite SOC, the spin group theory reduces to magnetic group theory. As indicated previously, the magnetic point group of spin canted $\rm MnSe$ becomes $2'$, leaving only $C_{2x}\mathcal{T}$ operator. The allowed LEE and NLEE components are summarized in Table~\ref{tab:selrules}, as denoted by the $\diamond$ symbol.

\begin{figure}[t]
\centering
\includegraphics[width=1.0\columnwidth]{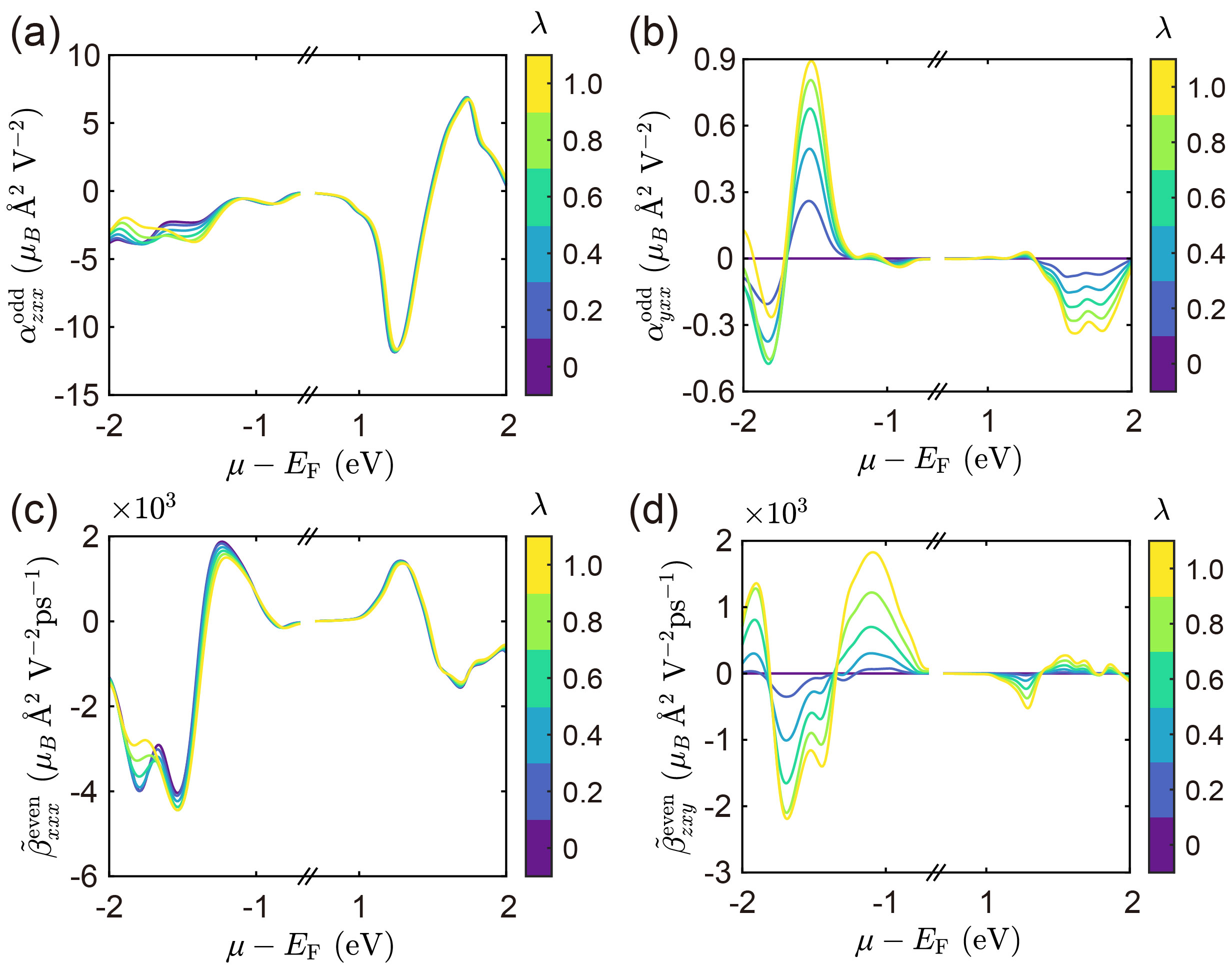}
\caption{Spin-orbit coupling strength $\lambda$ dependent on the (a) exchange field-induced $\alpha^{\rm odd}_{zxx}$, (b) SOC-induced $\alpha^{\rm odd}_{yxx}$, (c) exchange field-induced $\tilde{\beta}_{xxx}^{\rm even}$, and (d) SOC-induced $\tilde{\beta}_{zxy}^{\rm even}$. Here, $\lambda=0$ refers SOC-free limit, and $\lambda=1$ denotes full SOC. One sees that the (a) and (c) remains almost unchanged with $\lambda$, while (b) and (d) increases linearly with $\lambda$.}
\label{fig:soc}
\end{figure}

In this regard, we perform calculations with SOC included self-consistently in the first-principles calculations, for a typical $\kappa_y=0.34$. From Table \ref{tab:selrules}, we select representative NLEE components that are induced by SOC on the same time-reversal constraints with those already exist without SOC. Hence, we focus on $\alpha_{yxx}^{\rm odd}$ and $\tilde{\beta}_{zxy}^{\rm even}$ in this section. We plot their chemical potential dependence in Fig.~\ref{fig:soc}. Here, the parameter $\lambda$ denotes the SOC strength scale, with $\lambda=1$ indicating full SOC. One sees that the chirality-driven $\alpha_{zxx}^{\rm odd}$ [Fig.~\ref{fig:soc}(a)] almost remains its magnitude under SOC. On the other hand, the SOC-induced spin-NLEE $\alpha_{yxx}^{\rm odd}$ linearly increases with $\lambda$ [Fig.~\ref{fig:soc}(b)], reaching $0.9\,\mu_B\angstrom^2/\mathrm{V}^{2}$ at the full SOC limit. This is about one order of magnitude smaller than $\alpha_{zxx}^{\rm odd}$ . For the orbital-NLEE responses, the chiral spin induced $\tilde{\beta}_{xxx}^{\rm even}$ changes marginally with SOC strength [Fig.~\ref{fig:soc}(c)]. The SOC-induced $\tilde{\beta}_{zxy}^{\rm even}$ arises with $\lambda$ and reaches $2.2\times10^{3}\,\mu_B\angstrom^2/(\mathrm{V}^{2}\cdot\mathrm{ps})$. Note that in the quasi-2D material, the in-plane OAM that composes cycloid rotation along the confined $z$-direction is in general much smaller than the out-of-plane component. The SOC effect on orbital-NLEE should be smaller than the chiral-spin origin. This is consistent with the symmetry argument results as described previously, and the energy scale argument between the NR exchange field and SOC, in line with the SOC-expansion mechanism for response functions such as Hall effect transport \cite{Liu2025PRX}. In addition, from Table \ref{tab:selrules}, we note that each electric field component (both LEE and NLEE) only permits one type of SAM and OAM accumulations, which can be used to distinguish them in potential experiments. Nonetheless, here we mainly focus on the NR-limit Edelstein effects engineered by vector spin chirality, and the SOC effect is beyond the scope of the current study, and will be discussed in future works.

\section{Conclusion}
In conclusion, we have shown that vector spin chirality could switch on the nonlinear Edelstein responses in quasi-2D antiferromagnets without the need of SOC effect, in contrast to conventional SOC-driven torque mechanisms. We perform a spin group-theoretical analysis to showcase how the SAM and OAM accumulations arise in buckled quasi-2D honeycomb lattice under an in-plane electric field. These responses originate from ordinary and Zeeman-like band geometry quantities of the Bloch wavefunction. First-principles calculations on $\rm MnSe$ monolayer suggest that our calculated SOC-free NLEE is well within experimentally accessible regimes. Our results identify that the noncollinear spin structure can provide a general mechanism for sizable charge-to-magnetization conversion in the absence of SOC, with potential applicability to a broad range of chiral magnetic systems, including skyrmions, bi-merons, and kagom\'e systems. Moreover, our work provides a pathway toward probing band-geometric properties in the nonrelativistic regime.

\begin{acknowledgments}
\textit{Acknowledgments.} This work is supported by the National Natural Science Foundation of China (NSFC) under Grant Nos. 12374065. The authors thank the support of the 111 Project (B25007). The HPC platform of Xi'an Jiaotong University is also acknowledged.
\end{acknowledgments}

\textit{Data availability.} The data that support the findings of this article are not publicly available yet. The data are available from the authors upon reasonable request.

\bibliography{refs}

\end{document}